\documentclass[showpacs,amsmath,amssymb, prl,twocolumn]{revtex4}
\usepackage{graphicx}
\usepackage{dcolumn}
\usepackage{bm}

\begin{document}

\title{Super-magnetoresistance effect in triplet spin valves}

\author{F. Romeo$^{1,2}$ and R. Citro$^{1,2}$}
\affiliation{$^{1}$Dipartimento di Fisica "E.R. Caianiello", Universit\`a  di Salerno, I-84084 Fisciano (SA), Italy\\
$^{2}$CNR-SPIN Salerno, I-84084 Fisciano (SA), Italy}

\date{\today}
\begin{abstract}
We study a triplet spin valve obtained by intercalating a triplet superconductor \textit{spacer} between two ferromagnetic regions with non-collinear magnetizations. We demonstrate that the magnetoresitance of the triplet spin valve depends on the relative orientations of the \textbf{d}-vector, characterizing the superconducting state, and the magnetization directions of the ferromagnetic layers. For devices characterized by a long superconductor, the \textit{Cooper pairs spintronics} regime is reached allowing to observe the properties of a polarized current sustained by Cooper pairs only. In this regime a \textit{super-magnetoresistance effect}  emerges, and the chiral symmetry of the order parameter of the superconducting spacer is easily recognized.
 Our findings open new perspectives in designing devices based on the cooperative nature of ferromagnetic and triplet correlations in a spintronic framework.
\end{abstract}


\maketitle
\textit{Introduction}. The realization of devices based on synthetic materials offers interesting technological opportunities.
Within this class of devices, the \textit{spin valves} (SVs) are a notable example.
They are made of two ferromagnetic layers separated by a non-magnetic \textit{spacer} and show
a significant change in the electrical resistance depending on whether the magnetizations of ferromagnetic (FM) regions are in a parallel or an anti-parallel configuration. This effect, more evident in multilayer structures displaying the giant
magnetoresistance effect \cite{baibich_1988,hirota_2002,parkin_1993}, finds application in the information technology industry (e.g. sensors for hard disk drives,  magnetic-random-access memory).

The spacer properties strongly affect the SV response and, thus, various SV devices
have been proposed to study the phase coherent transport through spacers made of nanowires \cite{nanowire_SV},
carbon-nanotubes \cite{cnt_SV}, ballistic low-dimensional systems \cite{graphene_SV} and singlet-superconducting regions \cite{linder_mr09singlet,scth2011}.
All these devices, however, contain a \textit{scalar spacer}, i.e. a middle layer unable to add (non-magnetic) vectorial quantities relevant in determining the magnetoresistive response of the SV.

Due to the non-antagonist behavior in the presence of magnetism, triplet superconductors (TSCs) are natural candidates to study SVs having spacers with exotic (i.e. \textit{non-scalar}) properties.
%=============================================================================fig1============================================================================
\begin{figure}[!t]
\includegraphics[clip,scale=0.27, angle=-90]{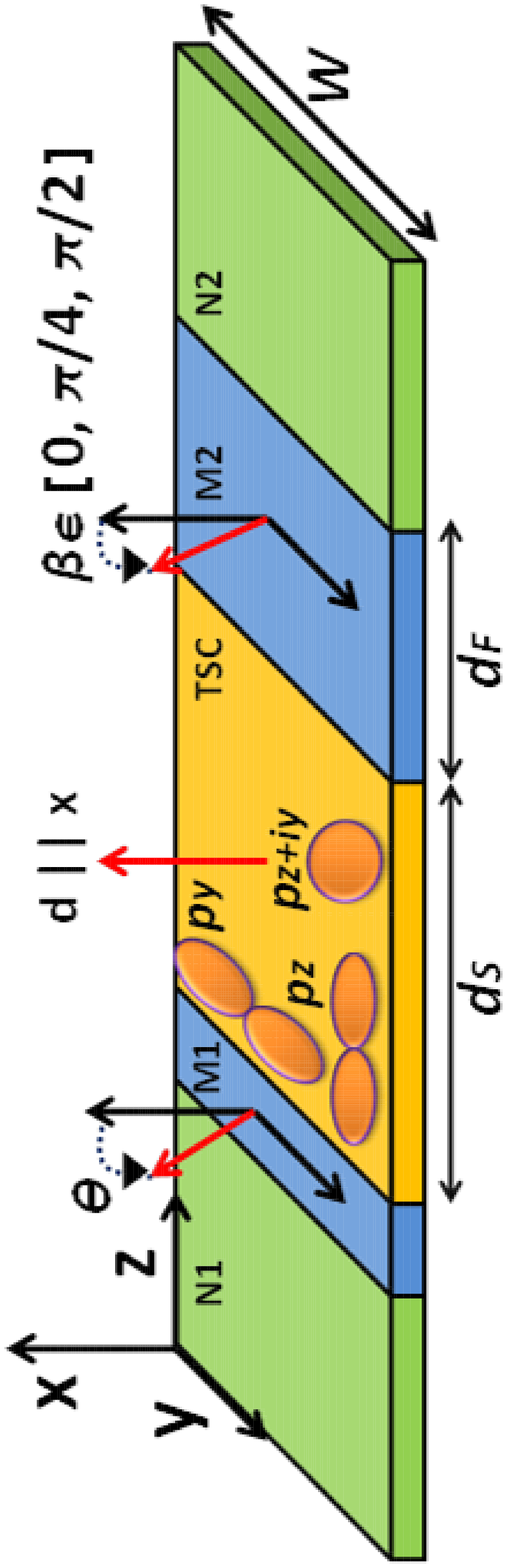}\\
\includegraphics[clip,scale=0.5]{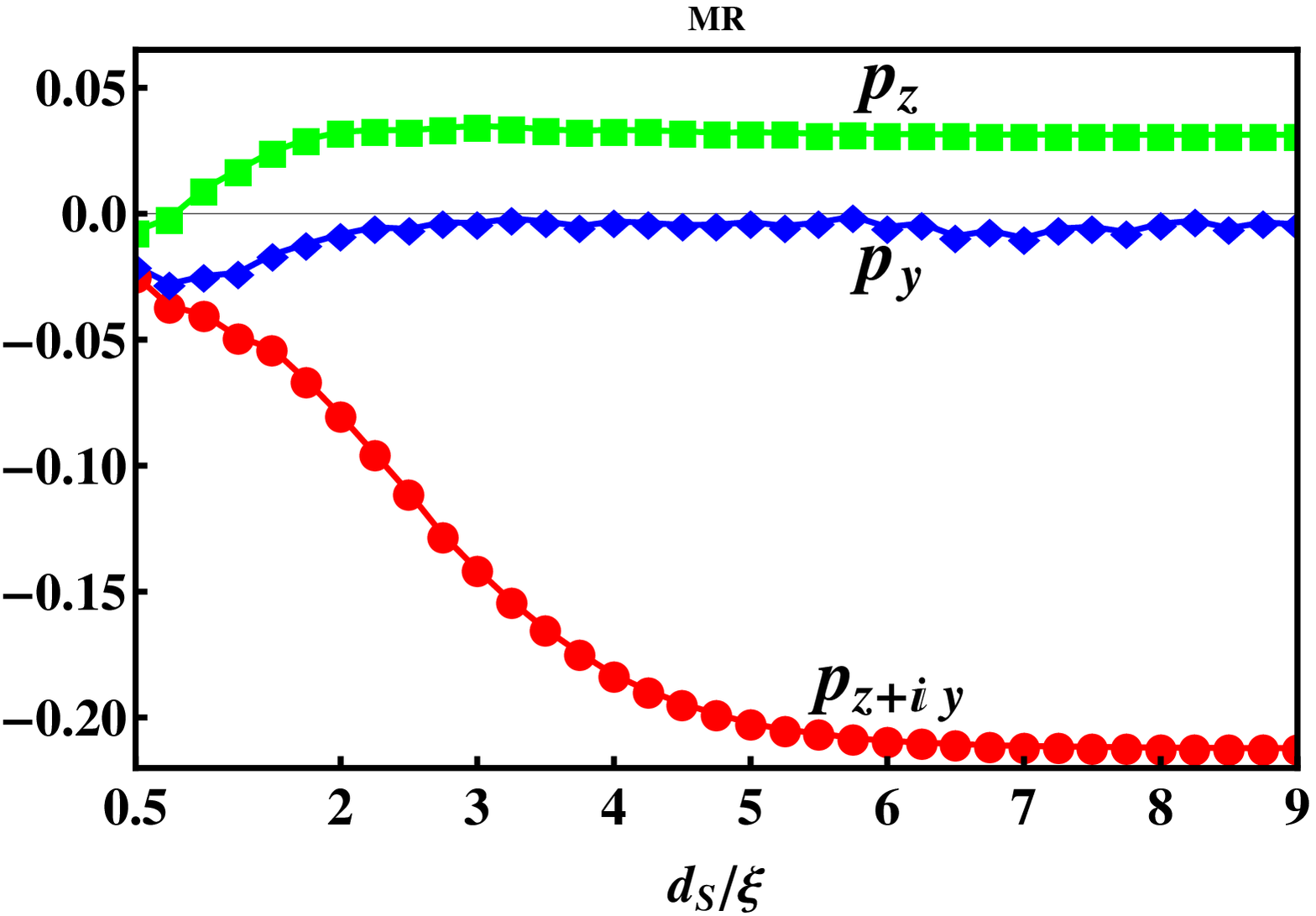}\\
\caption{(Color online) (Upper panel) Triplet-spin-valve device. A triplet superconductor (\textbf{TSC}) is intercalated between the nanostructured magnetic regions \textbf{M1} and \textbf{M2} whose magnetic momenta $\vec{\mathrm{M}}_1$ and $\vec{\mathrm{M}}_2$, belonging to the $x-y$ plane, form the angle $\theta$ and $\beta \in [0,\pi/4,\pi/2]$ with respect to the \textbf{d}-vector (parallel to the $x$ direction) characterizing the superconducting state. The system is biased by means of non-magnetic leads \textbf{N1} and \textbf{N2} having a transverse dimension $W$. (Lower panel) Magnetoresitance curves as a function of the superconductor length $d_S$. The model parameters are fixed as follows: $\varepsilon/\Delta=0.01$, $\mathbf{h}=0.65$, $\Gamma=1.5$, $Z_{BTK}=1$, $\theta=\pi/2$, $\beta=\pi/4$, $d_F=\xi/10$. Differently from an s-wave spin valve, the MR is a non-vanishing function of $d_S$.}
\label{fig:device}
\end{figure}
%===============================================================================================================================================================
Since the discovery of triplet superconductivity
in $\mathrm{Sr_2RuO_4}$~\cite{Maeno1994} there has been
growing interest in the properties of TSC heterostructures ~\cite{Barash2001,Asano,Kastening2006,Tanuma2006,Norway,BrydonTFT2008,Brydonlett2009,Kuboki2004,Cuoco2008,Tanaka2007}.
Despite this,  the study of devices combining TSCs and ferromagnets  is
still in its infancy and only few unconventional effects have been predicted~\cite{Kastening2006,BrydonTFT2008,Brydonlett2009,Brydon2009}.

In this Letter, we study the magnetoresistance (MR) properties of a SV  whose spacer is a triplet superconductor having the vector order parameter, the so-called \textbf{d}-vector, parallel to the $x$-direction (see upper panel, Fig.\ref{fig:device}). This system is the prototype of a SV having a \textit{vectorial spacer} which has never been  discussed before. We demonstrate that triplet superconducting spin valves can transmit information by means of dissipationless Cooper pairs current, showing a symmetry-dependent (i.e. $p_z$, $p_y$ or $p_{z+iy}$) \textit{super-magnetoresistance} effect. Due to the presence of a \textit{vectorial spacer} (i.e. the TSC), the magnetoresistive response of the device does not exhibit a Jullier-like \cite{jullier_formula} behavior, being the MR affected by the relative orientation of the \textbf{d}-vector with respect to the polarizations $\vec{\mathrm{M}}_1$, $\vec{\mathrm{M}}_2$ of the FM regions.
Furthermore, differently from a singlet superconducting spacer, the MR is not vanishing as a function of $d_s$, the length of the superconducting layer (see lower panel, Fig.\ref{fig:device}), and the triplet correlations induce a considerable MR for spacer length $d_s>4 \xi$, $\xi$ being the superconducting coherence length. This is a specially distinguishable feature of the chiral symmetry of the order parameter.

\textit{Theoretical description}. A triplet SV (TSV) consists of two ferromagnetic regions, namely M1 and M2, separated by a spacer realized by a TSC (Fig.\ref{fig:device}). The whole system is connected to semi-infinite non-magnetic leads N1 and N2, while translational invariance along the $y$-direction is assumed ($W \gg\lambda_F$, $\lambda_F$ being the Fermi wavelength). The Bogoliubov-de Gennes (BdG) equation describing the quasiparticle states with energy $E$ is written in Nambu notation as ($r\equiv(y,z)$)
\begin{equation}
\label{eq:hamiltonian}
\left[
  \begin{array}{cc}
    \hat{H}(r) & \hat{\Delta}(r) \\
    \hat{\Delta}^{\dagger}(r) & -\hat{H}^{\ast}(r) \\
  \end{array}
\right]\Psi(r)=E\Psi(r)
\end{equation}
where the hat sign indicates $2 \times 2$ matrices in spin-space. The single particle Hamiltonian can be written as follows:
\begin{equation}
\hat{H}(r)=\Bigl[-\frac{\hbar^2\nabla^2}{2m}-E_F+V_{int}(r)\Bigl]\hat{\mathbf{1}}-g\mu_B\mathbf{\hat{\sigma}} \cdot\mathbf{M}(r),
\end{equation}
where we introduced an interface potential controlling the barrier transparencies $V_{int}(r)=U[\delta(z)+\delta(z-d_S)+\delta(z-d_S-d_F)]$ and the local magnetic fields describing the ferromagnetic regions M1 and M2:
\begin{eqnarray}
\mathbf{M}(r)&=&M_1 \delta(z) \mathbf{v}(\theta)+M_2(r) \mathbf{v}(\beta)\\
\mathbf{v}(\phi)&=&\cos(\phi)\mathbf{e}_x+\sin(\phi)\mathbf{e}_y,
\end{eqnarray}
$\mathbf{e}_{x/y/z}$ being the orthogonal triad of unit vectors.
The region M1, corresponding to the so-called free-layer in the SV language, is assumed to be very narrow and thus is modeled using a Dirac delta potential whose amplitude is proportional to the magnetic momentum $\mathrm{M}_1$ of the layer. The function $\mathrm{M}_2(r)$, describing the magnetization of the region M2, is taken spatially homogeneous inside the region and zero elsewhere.\\
The gap matrix in Eq.\ref{eq:hamiltonian} is defined as $\hat{\Delta}(r)=i[\hat{\mathbf{\sigma}} \cdot \mathbf{d}(r)]\hat{\sigma}_y$, where $\mathbf{d}(r)$ is the vector defining the order parameter of the TSC. Here we are interested in describing superconducting spacers exhibiting equal-spin-pairing unitary states for which $\mathbf{d}(r)=\Delta(r)\mathbf{e}_x$ represents a convenient choice. With this assumption, the triplet Cooper pairs have $z$-component of the spin $S_z=\pm \hbar$, while the condensate does not present a net spin polarization. The magnitude of the gap $\Delta$ is assumed to be constant throughout the superconducting region and zero elsewhere. In the following we consider three orbital pairing states: $p_y$-wave, $\Delta_k=\Delta k_y/k_F$; $p_z$-wave, $\Delta_k=\Delta k_z/k_F$; and the chiral $p_{z+iy}$-wave, $\Delta_k=\Delta [k_{z}+i k_y]/k_F$ \cite{note0}.

The translational invariance along the $y$-direction implies the conservation of the wave-vector $k_y$ parallel to the interface. As a consequence, the wavefunction can be written as $\Psi(r)=e^{ik_y y}\psi(z|E,k_y)$ leading to an effective one-dimensional scattering problem for the wavefunction $\psi(z|E,k_y)$, being the energy $E$ and $k_y$ conserved quantum numbers in the scattering events. Once the scattering wavefunctions have been written in the regions $\mathrm{N}_{1,2}$, TSC and M2, the scattering matrix $\mathcal{S}$, describing the transport properties, is obtained by solving the system of equations imposed by the boundary conditions (BCs) at the interfaces. For instance, the discontinuity at $z=0$ described by the local potential $[U-g\mu_B \mathrm{M}_1\hat{\sigma}\cdot \mathbf{v}(\theta)]\delta(z)$ implies the following BCs: (i) $\psi(z=0^{+}|E, k_y)=\psi(z=0^{-}|E, k_y)$; (ii) $\partial_z\psi(z=0^{+}|E, k_y)-\partial_z\psi(z=0^{-}|E, k_y)=k_F[1_{4 \times 4}Z_{BTK}-\Gamma \mathcal{A}(\theta)]\psi(z=0^{+}|E, k_y)$, where we have introduced
\begin{equation}
\mathcal{A}(\theta)=\left[
                      \begin{array}{cc}
                        \hat{\sigma}\cdot \mathbf{v}(\theta) & 0 \\
                        0 & \hat{\sigma}^{\ast}\cdot \mathbf{v}(\theta) \\
                      \end{array}
                    \right],
\end{equation}
the BTK parameter \cite{btk} $Z_{BTK}=2mU/(\hbar^2 k_F)$ and its generalization to the spin-active barrier case $\Gamma=2m[g\mu_B \mathrm{M}_1]/(\hbar^2 k_F)$. As a consequence the $\mathcal{S}$-matrix depends on the incidence angle $\alpha_{in}$  through the conserved quantity $k_y=|k(E)|\sin(\alpha_{in})$. The conductance of the system can be obtained within the scattering field theory \cite{buttiker92} by defining the field
 \begin{eqnarray}
 \hat{\Psi}_j(z,t|k_y)&=&\sum_{\beta,\sigma} \int\frac{dE e^{-iEt}}{\sqrt{2\pi\hbar |v_z(E)|}}|\beta\rangle\otimes|\sigma\rangle\times\\\nonumber
 &&[\hat{a}^{\sigma}_{j\beta}(E;k_y)e^{ik^{(z)}_{\beta}z}+\hat{b}^{\sigma}_{j\beta}(E;k_y)e^{-ik^{(z)}_{\beta}z}]
 \end{eqnarray}
and by evaluating the $k_y$-resolved conductance tensor elements $g_{ik}(k_y)$ according to the derivation given in Ref. \cite{scth2011}. The scattering operators $\hat{a}^{\sigma}_{j\beta}(E;k_y)$ ($\hat{b}^{\sigma}_{j\beta}(E;k_y)$) destroy an incoming (outgoing) particle of species $\beta \in \{e,h\}$ and spin projection $\sigma\in\{\uparrow,\downarrow\}$ in the lead $j$, while the wave vector $k^{(z)}_{\beta}=\eta_{\beta}k_z(E)=\eta_{\beta}|k(E)|\cos(\alpha_{in})$ ($\eta_e=-\eta_h=1$) along the transport direction is written within the Andreev approximation \cite{note}, i.e. $v^{j\beta\sigma}_z(E)\approx v_z(E)=\hbar |k(E)|\cos(\alpha_{in})/m$. The scattering matrix relates the incoming and the outgoing processes according to the equation:
\begin{equation}
\hat{b}^{\sigma}_{j\beta}(E;k_y)=\sum_{\beta'\sigma'j'} \mathcal{S}^{\beta\beta'}_{jj'\sigma\sigma'}(E;k_y)\hat{a}^{\sigma'}_{j'\beta'}(E;k_y),
\end{equation}
which is sensitive to the properties of the scattering region.
%====================================================================================================
\begin{figure*}[!t]
\resizebox{2.0\columnwidth}{!}{%
  \includegraphics[clip,scale=0.5]{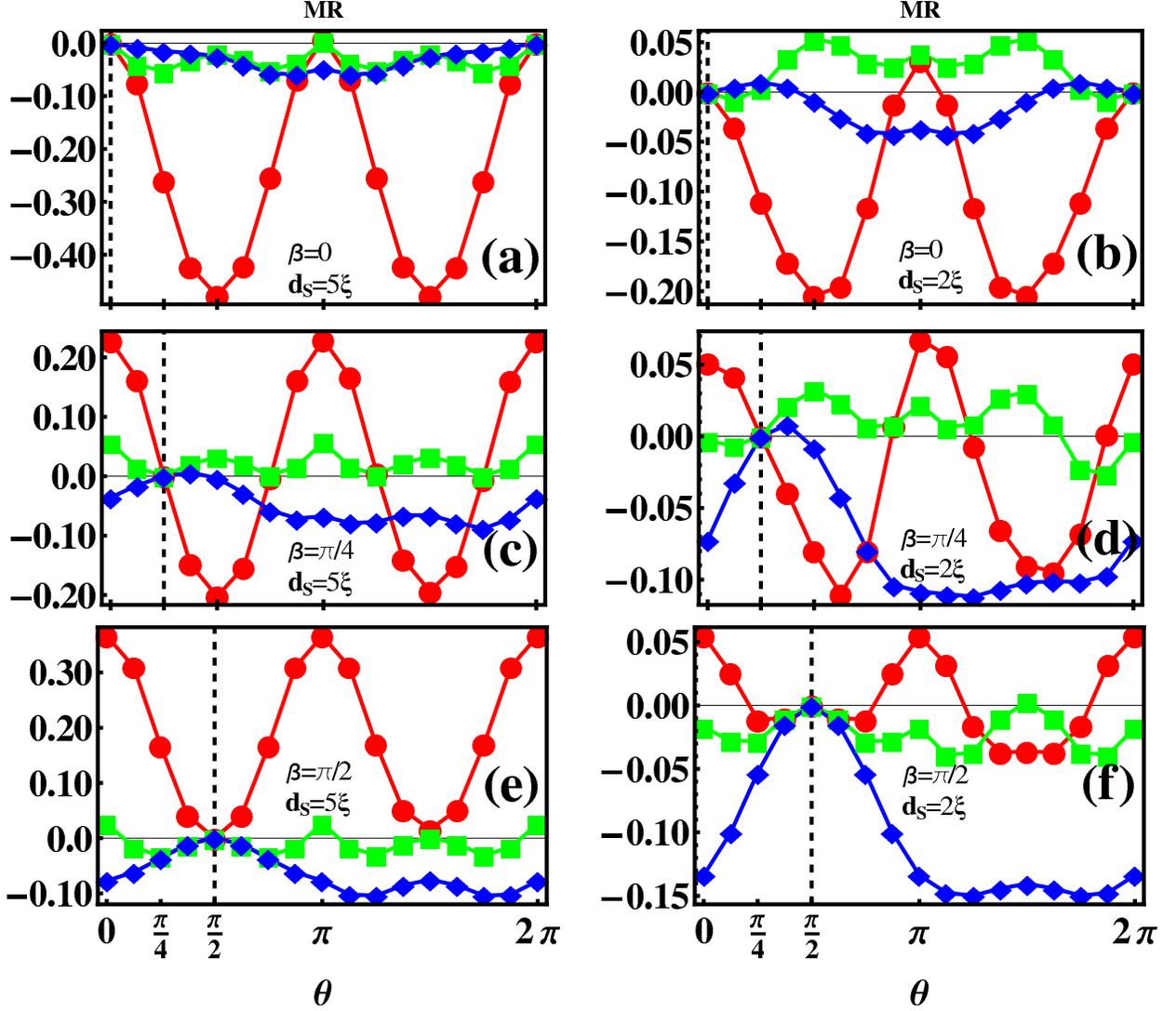}
} \caption{(Color online) Magnetoresistance curves as a function of the angle $\theta$ computed for the order parameter symmetry $p_z$ (green box $\blacksquare$), $p_y$ (blue diamond $\blacklozenge$), $p_{z+iy}$ (red circle $\bullet$). The curves \textbf{(a)}, \textbf{(c)}, \textbf{(e)} are computed for $d_S=5\xi$, while the curves \textbf{(b)}, \textbf{(d)}, \textbf{(f)} are computed by setting $d_S=2\xi$. The magnetization of the region M2 is fixed to form the angle: $\beta=0$ with the \textbf{d}-vector for the panels \textbf{(a)}-\textbf{(b)}; $\beta=\pi/4$ for the panels \textbf{(c)}-\textbf{(d)};  $\beta=\pi/2$ for the panels \textbf{(e)}-\textbf{(f)} (see the vertical dashed line). The remaining model parameters are fixed as follows: $\varepsilon/\Delta=0.01$, $\mathbf{h}=0.65$, $\Gamma=1.5$, $d_F=\xi/10$, $Z_{BTK}=1$. The figures show that when $d_S$ becomes larger than a symmetry-dependent length the $\mathrm{MR}(\theta,\beta)$ curves show a period halving consisting of a transition from $2\pi$ periodicity (for $d_S=2\xi$) to $\pi$ (for $d_S=5\xi$). The latter transition is not observed for the $p_y$ symmetry whose periodicity is always $2\pi$.}
\label{fig2}
\end{figure*}
%===============================================================================================================
The linear response current $\mathcal{I}_i$ flowing through the i-th lead is given by the sum of independent contributions of the elementary processes labeled by all possible $k_y$, i.e. $\mathcal{I}_i=\sum_{j,k_y}g_{ij}(k_y)(\mu_j-\mu_s)$; $\mu_j$ is the electrochemical potential of the j-th lead while in the superconducting region $\mu_s$ is fixed by imposing the electric charge conservation. The two-terminal conductance is thus given by $\mathcal{G}=[\bar{g}_{22}\bar{g}_{11}-\bar{g}_{21}\bar{g}_{12}]/(\sum_{ij}\bar{g}_{ij})$ with the definition $\bar{g}_{ij}=\sum_{k_y}g_{ij}(k_y)$ \cite{note1}. In the symmetric case the conductance $\mathcal{G}$ takes the simplified form $\mathcal{G}=(\bar{g}_{11}-\bar{g}_{12})/2$ which can be explicitly written in terms of the scattering matrix elements as follows \cite{note2}:
\begin{eqnarray}
&&\mathcal{G}=\frac{e^2 k_F W}{\pi h}\int d\xi d\alpha_{in}[-\partial_{\xi}f(\xi)]_{eq} \Bigl[\frac{\cos(\alpha_{in})}{2}\Bigl]\times\\\nonumber
&&[\sum_{\beta\in\{e,h\}}\mathcal{M}^{\beta\beta}_{12}(\xi,\alpha_{in})+\mathcal{M}^{he}_{11}(\xi,\alpha_{in})+\mathcal{M}^{eh}_{11}(\xi,\alpha_{in})],
\end{eqnarray}
where $\mathcal{M}^{\alpha\beta}_{ij}=Tr[\mathcal{S}_{ij}^{\alpha\beta\dag}\mathcal{S}_{ij}^{\alpha\beta}]$, being the trace performed over the spin degree of freedom. When the incidence angle $\alpha_{in}$ is constrained to the interval $[-\zeta,\zeta]$, the sum rule $\sum_{j\alpha; k_y}\mathcal{M}^{\beta\alpha}_{ij}(E,k_y)=2\mathcal{N}_{\bot}$ is obeyed with the transverse modes number given by $\mathcal{N}_{\bot}=k_FW\sin(\zeta)/\pi$.\\
\textit{Results}. A TSV device is a spintronic system where the competition between conventional Andreev reflection (CAR) and unconventional Andreev reflection (UAR) induce specific features in the magnetoresistance curves.
Here, we define the magnetoresistance as $\mathrm{MR}=1-\mathcal{G}(\theta,\beta)\mathcal{G}(\theta=\beta,\beta)^{-1}$, being $\mathcal{G}(\theta=\beta,\beta)$ the conductance in parallel configuration of the magnetic momenta of the ferromagnetic regions. This quantity is a function of the angle $\beta$ formed by M1/M2 with the \textbf{d}-vector. The Andreev reflection (AR) processes, dominating the sub-gap transport, are strongly affected by $\beta$ and $\textbf{h}$, the Zeeman energy, and the probabilities of these elementary processes present a complicated behavior. In the presence of a spin active interface, two elementary processes are possible: (i) the conventional AR $(e,\sigma)_{in}\rightarrow(h,\sigma)_{out}$; (ii) the unconventional AR $(e,\sigma)_{in}\rightarrow(h,\bar{\sigma})_{out}$, where $\bar{\sigma}$ represents the spin polarization opposite to $\sigma$. The probabilities of the conventional AR, $\mathcal{P}_{CAR}$, and of the unconventional AR, $\mathcal{P}_{UAR}$, are comparable for $\beta=\pi/4$, while $\mathcal{P}_{UAR}>\mathcal{P}_{CAR}$ ($\mathcal{P}_{UAR}<\mathcal{P}_{CAR}$) for $\beta \in [0,\pi/4]$ ($\beta \in [\pi/4,\pi/2]$). From the physical viewpoint a relevant $\mathcal{P}_{UAR}$ is associated to the pair breaking effect exerted by an out-of-plane component (parallel to the $x$-direction) of the magnetic momentum of M2 on the in-plane ($yz$) spin polarization of the Cooper pairs.\\
The effects of the interplay discussed above is studied within the zero temperature limit ($T=0$) which is appropriate to study SV containing TSCs whose typical critical temperatures $T_c$ are quite low. For instance in $\mathrm{Sr_2RuO_4}$, we have $T_c\approx 0.7-1.4 \mathrm{K}$ and the zero temperature coherence length $\xi(T=0)\approx 70\mathrm{nm}$ \cite{sro}. In the computation we set the maximal incidence angle $\zeta=35^{\circ}$ to mimic the angular dispersion of electron waves coming from a remote constriction. The ferromagnetic region M2 is made of a weak-ferromagnet characterized by normalized Zeeman energy $\mathbf{h}=(g\mu_BM_2)/E_F$ such that $\mathbf{h} \in[0.05, 0.65]$, while the spin-dependent wavevector $k^{\sigma}_z\approx k_F\sqrt{\cos^{2}(\alpha_{in})+\sigma \mathbf{h}}$ is a real quantity for both spin polarizations.\\
In Figs. \ref{fig2} the magnetoresistance curves as a function of the angle $\theta$ for short ($d_S=2\xi$, for Figs. \ref{fig2}(\textbf{b}), \ref{fig2}(\textbf{d}), \ref{fig2}(\textbf{f})) and long ($d_S=5\xi$, for Figs. \ref{fig2}(\textbf{a}), \ref{fig2}(\textbf{c}), \ref{fig2}(\textbf{e})) spacers are shown. The magnetization direction of the region M2 is fixed to $\beta=0$ (Figs. \ref{fig2}(\textbf{a})-(\textbf{b})), $\beta=\pi/4$ (Figs. \ref{fig2}(\textbf{c})-(\textbf{d})) and $\beta=\pi/2$ (Figs. \ref{fig2}(\textbf{e})-(\textbf{f})).
The magnetoresistance curves pertaining to the symmetries $p_z$ and $p_{z+iy}$ shown in Fig. 2(\textbf{a}), 2(\textbf{c}), 2(\textbf{e}) are described by the relation $\mathrm{MR}(\theta, \beta)=\mathcal{F}(\theta)-\mathcal{F}(\beta)$, where $\mathcal{F}(x)=\sum_{n=1,2}\mathcal{B}_n \cos^{n}(2x)$ is a symmetry-dependent function. The $p_{z+iy}$ symmetry is well described by the function $\mathcal{F}(x)$ characterized by $\mathcal{B}_1 \approx0.2 $ and $\mathcal{B}_2=0$; the case of $p_{z}$ is instead described by $\mathcal{B}_1 \approx (0.6-0.9)\cdot 10^{-2}$ and $\mathcal{B}_2/ \mathcal{B}_1\approx 5-9$. For both cases $\mathrm{MR}(\theta, \beta)$ is a separable function of $\theta$ and $\beta$, while for short superconducting regions (see Figs. 2(\textbf{b}), 2(\textbf{d}), 2(\textbf{f}) for $d_S=2\xi$) the magnetoresistance curves show a complicated behavior induced by interference effects assisted by partial conversion of the quasiparticles current into Cooper pairs current.\\
The magnetoresistance curves of the $p_y$-symmetry case present a periodicity of $2\pi$ with respect to $\theta$ for all the regimes considered here. This phenomenon is related to the existence of a gapless line ($\alpha_{in}=0$) along the transport direction. Under this condition, a quasiparticles flux always  coexists with a Cooper pair current. The two fluxes contribute differently to the periodicity of the magentoresistance: the Cooper pairs flux induces a $\pi$-periodicity, while the quasiparticles flux presents a $2\pi$-behavior. The global periodicity is dominated by the longer period and thus a $\pi$-periodicity is never observed as long as a non-vanishing quasiparticles flux contributes to the conductance. This observation explains the period halving of the magnetoresistance curves for the symmetries $p_z$ and $p_{z+iy}$.
\begin{figure}[!h]
\includegraphics[clip,scale=0.85]{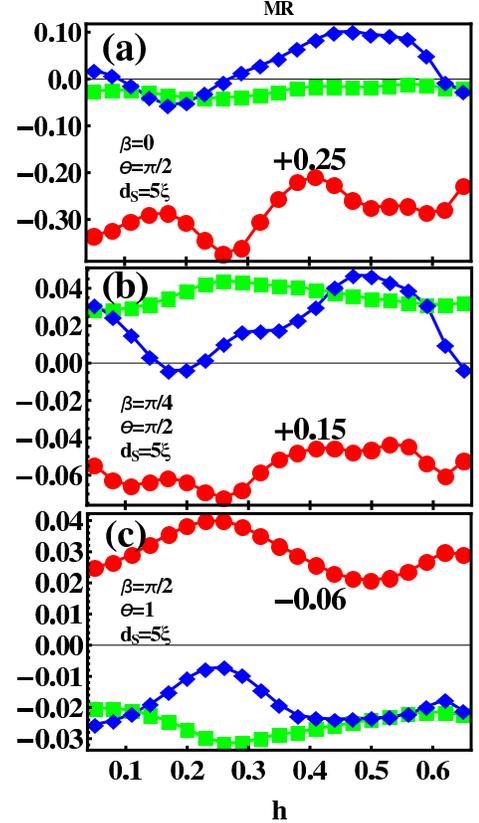}
\caption{Magnetoresistance curves as a function of the Zeeman energy $\mathbf{h}$ of the region M2 computed for the order parameter symmetry $p_z$ (green box $\blacksquare$), $p_y$ (blue diamond $\blacklozenge$), $p_{z+iy}$ (red circle $\bullet$).  The magnetization of the region M2 is fixed to form the angle: $\beta=0$ with the \textbf{d}-vector (panel \textbf{(a)}); $\beta=\pi/4$ (panel \textbf{(b)});  $\beta=\pi/2$ (panel \textbf{(c)}). The remaining model parameters are fixed as follows: $d_S=5\xi$, $\varepsilon/\Delta=0.01$, $\Gamma=1.5$, $d_F=\xi/10$, $Z_{BTK}=1$. The value $\theta=\pi/2$ has been used to compute the curves shown in the panels (\textbf{a})-(\textbf{b}), while $\theta=1$ has been used in the panel (\textbf{c}). All the magnetoresistance curves pertaining to the chiral symmetry $p_{z+iy}$ have been shifted by adding a constant offset written close to the curves.}
\label{fig:h}
\end{figure}
For short channel devices ($d_S=2\xi$), the quasiparticles current significantly contributes to the conductance determining a dominant $2\pi$-periodicity. On the other hand, by increasing the length of the superconducting spacer up to $d_S=5\xi$ a bulk-like behavior, characterized by a dissipationless Cooper pairs current, is established and a $\pi$-periodicity is observed. The latter regime is not observable for SV having a singlet superconducting spacer where, provided that the magnetoresistance is non-vanishing, the polarized transport is always sustained by a quasiparticles current. Under this condition, the Jullier-like behavior $\mathrm{MR}(\theta, \beta) \propto \cos(\theta-\beta)$ is obeyed \cite{jullier_formula}.\\
An additional fingerprint of the triplet correlations is represented by the high magnetoresistance values compared to the s-wave case \cite{scth2011}. In the low-transparency regime ($Z_{BTK}=1$) studied in this Letter, the chiral symmetry spacer ($p_{z+iy}$) induces higher magnetoresistance values compared to the ones obtained for $p_z$ and $p_y$-symmetry spacers. The latter behavior is not modified by varying the modulus of polarization $\textbf{h}$ of the region M2. This aspect is visible in Figs.\ref{fig:h}, where the MR curves as a function of $\textbf{h}$ are shown, assuming $d_S=5\xi$ and considering the different magnetization directions of M2, i.e. $\beta\in\{0, \pi/4, \pi/2\}$. In all the panels, the chiral spacer shows higher MR values compared to the other symmetries, while an oscillating behavior as a function of $\mathbf{h}$ is observed. The oscillations can induce a sign reversal of the MR curves for the symmetries $p_y$, while the remaining symmetries do not change sign by the variation of $\mathbf{h}$. For shorter systems characterized by $d_S=2\xi$ (not shown here) and for appropriate magnetic configurations of the region M1/M2 a sign reversal is also possible for the chiral and the $p_z$ case. This behavior originates by the interplay of spin-sensitive interference effects of quasiparticles in the scattering region coexisting with conventional and anomalous Andreev processes.\\
\textit{Conclusions}. In conclusion, a SV modified by the inclusion of a triplet superconducting spacer is a novel system of spintronic  relevance, displaying unconventional magnetoresistive response. Differently from SVs with normal or s-wave spacers, a TSV shows a magnetoresistive behavior which depends on the relative orientations of the three vectors $\mathrm{M}_1$, $\mathrm{M}_2$ and the \textbf{d}-vector. A non-vanishing symmetry-dependent MR, \textit{super-magnetoresistance}, has been obtained for long spacers ($d_S \approx 5\xi$) allowing the study of spintronic properties completely determined by the dissipationless spin polarized currents sustained by the Cooper pairs. Experimentally, the regime of pure \textit{Cooper pairs spintronics} is signaled by a $\pi$-periodicity of the MR \textit{vs} $\theta$ curves for the chiral and the $p_z$ symmetries; a $2\pi$-periodicity is always found in the $p_y$-case. These findings open new perspectives in designing new experiments aiming to study the cooperative nature of ferromagnetic and triplet correlations in a spintronic framework.\\
\textit{Acknowledgements}.
The authors would like to thank F. Giubileo and M. Cuoco for many useful discussions.


\begin{thebibliography}{99}
\bibitem{note0}The gap operator is defined according to $\hat{\Delta}(r)=i\hat{\sigma}_x\hat{\sigma}_y\Delta(r)$, $\Delta(r)$ being the differential operator: $\Delta(r)=\frac{\Delta}{\hbar k_F}\hat{p}_z$ ($\Delta(r)=\frac{\Delta}{\hbar k_F}\hat{p}_y$) for $p_z$-symmetry ($p_y$-symmetry); $\Delta(r)=\frac{\Delta}{\hbar k_F}[\hat{p}_z+i\hat{p}_y]$ for the chiral symmetry. $\Delta_k$ is defined by the action of $\Delta(r)$ on the plane wave $\phi(r)=\exp(ik_z z+ik_y y)$, according to the relation: $\Delta(r)\phi(r)=\Delta_k\phi(r)$.
\bibitem{note}The Andreev approximation is not performed in the scattering region to correctly describe the coherent transport inside the scattering region.
\bibitem{note1}The sum over the transverse uncoupled modes labeled by $k_y$ is performed according to the prescription: $\sum_{k_y}F(k_y)\rightarrow \frac{k_F W}{2\pi}\int_{[-\zeta,\zeta]}d\alpha_{in}\cos(\alpha_{in})F(k_F\sin(\alpha_{in}))$, $\zeta$ being the maximal incidence angle.
\end{thebibliography}
\end{document}